\documentclass[a4paper]{PoS}

\usepackage[nolist]{acronym}

\begin{acronym}
	\acro{ATLAS}{Australia Telescope Large Area Survey}
	\acro{EMU}{the Evolutionary Map of the Universe}
	\acro{ASKAP}{Australian Square Kilometre Array Pathfinder}
	\acro{SSDF}{Spitzer-South Pole Telescope Deep Field}
	\acro{BCS}{Blanco Cosmology Survey}
	\acro{DES}{Dark Energy Survey}
	\acro{VHS}{VISTA Hemisphere Survey}
	\acro{WISE}{Wide-field Infrared Survey Explorer}
	\acro{GLEAM}{The GaLactic and Extragalactic All-sky MWA}
	\acro{SPT}{South Pole Telescope}
	\acro{RFI}{Radio-frequency interference}
	\acro{PyBDSM}[\textsc{PyBDSM}]{Python Blob Detection and Source Measurement}
	\acro{AGN}{Active Galactic Nuclei}
	\acro{SZ}{Sunyaev-Zel'dovich}
	\acro{ATCA}{Australia Telescope Compact Array}
	\acro{ICM}{intra-cluster medium}
\end{acronym}

\def\deg{$^{\circ}$}
\def\sqdeg{~deg$^2$}
\def\arcsec{"}

\def\ujybm{~$\mu$Jy/beam}
\def\ujy{~$\mu$Jy}
\def\miriad{\textsc{miriad}}

\title{The ATLAS-SPT Radio Survey of Cluster Galaxies}

\ShortTitle{The ATLAS-SPT Survey}

\author{\speaker{A. N. O'Brien}$^{ab}$, N. F. H. Tothill$^{a}$, R. P. Norris$^{ab}$, M. D. Filipovi\'{c}$^{a}$\\
        \llap{$^{a}$}Western Sydney University, Penrith NSW, Australia\\
        \llap{$^{b}$}CSIRO Astronomy and Space Science, Epping NSW, Australia\\
        E-mail: \email{andrew.obrien@westernsydney.edu.au}}

\abstract{Using a high-performance computing cluster to mosaic 4,787 pointings, we have imaged the 100 sq. deg. South Pole Telescope (SPT) deep-field at 2.1 GHz using the Australian Telescope Compact Array to an rms of 80\ujy{} and a resolution of 8". Our goal is to generate an independent sample of radio-selected galaxy clusters to study how the radio properties compare with cluster properties at other wavelengths, over a wide range of redshifts in order to construct a timeline of their evolution out to $z \sim 1.3$. A preliminary analysis of the source catalogue suggests there is no spatial correlation between the clusters identified in the SPT-SZ catalogue and our wide-angle tail galaxies.}

\FullConference{EXTRA-RADSUR2015 (*)\\
		20--23 October 2015\\
		Bologna, Italy

                \bigskip
                \hrule
                \bigskip

                \textnormal{(*) This conference has been organized
                  with the support of the Ministry of Foreign Affairs
                  and International Cooperation, Directorate General
                  for the Country Promotion (Bilateral Grant Agreement
                  ZA14GR02 - Mapping the Universe on the Pathway to
                  SKA)}
}

\begin{document}

\section{Introduction}
We have conducted the next generation (ATLAS-SPT) of the \ac{ATLAS} \cite{Norris2006} to cover the 100\sqdeg{} \ac{SSDF} at a central frequency of 2.1 GHz, achieving a resolution of approximately 8\arcsec{} with a typical rms of 80\ujy. Due to the sensitivity, large area and available multi-wavelength coverage of the field, the science goals for this survey are broad. The primary goals are to conduct an evolution study of radio selected galaxy clusters and investigate radio source distributions to explore large-scale structure.

In the context of previous \ac{ATLAS} surveys and its planned successor \ac{EMU} \cite{Norris2011} using the \ac{ASKAP}, this survey (ATLAS-SPT) is a factor of 5 shallower than the deepest existing \ac{ATLAS} maps, however it is more than an order of magnitude larger in area. Where \ac{ATLAS} has so far detected approximately 6,000 galaxies, we expect to find about 20,000 in the final data products of ATLAS-SPT.

The \ac{SSDF} has been observed with a vast array of multi-wavelength instruments: the inner 25\sqdeg{} has been covered with XMM-Newton as part of the XXL project; optical data from the \ac{BCS} \cite{Desai2012} and images from the \ac{DES}; near-IR from the \ac{VHS}; mid-IR from \ac{WISE}; far-IR from Herschel-SPIRE; and low-frequency radio coverage from the \ac{GLEAM} survey. Our survey fills the large centimetre-wavelength radio gap in coverage. In addition to the aforementioned coverage, the field is also covered by the \ac{SPT} survey at 90, 150 and 220 GHz with an accompanying catalogue of \ac{SZ} detected galaxy clusters.

In this article we describe the survey parameters, data reduction, imaging process and preliminary results regarding the detection of galaxy cluster candidates using bent-tail radio galaxies.

\section{Survey Design}
ATLAS-SPT covers the the 100\sqdeg{} SSDF between 23h $\leq \alpha \leq$ 24h (0h) and -60\deg $\leq \delta \leq$ -50\deg at 1.1-3.1~GHz. The project (ATNF project code: C2788; PI: N. F. H. Tothill) was awarded 243 hours of observing time on the ATCA in the 2013 April semester, with an additional 17 hours awarded in the 2013 October semester to recover lost data due to interference. The ATCA receivers have linear feeds which record all Stokes parameters (XX, YY, XY, YX) however only total intensity (Stokes I) is used in this study. The observations were made with the array in the 6A and 6C configurations which produced images with a synthesised beam resolution of approximately 8\arcsec{}.

\begin{figure}
   \includegraphics[width=0.5\textwidth]{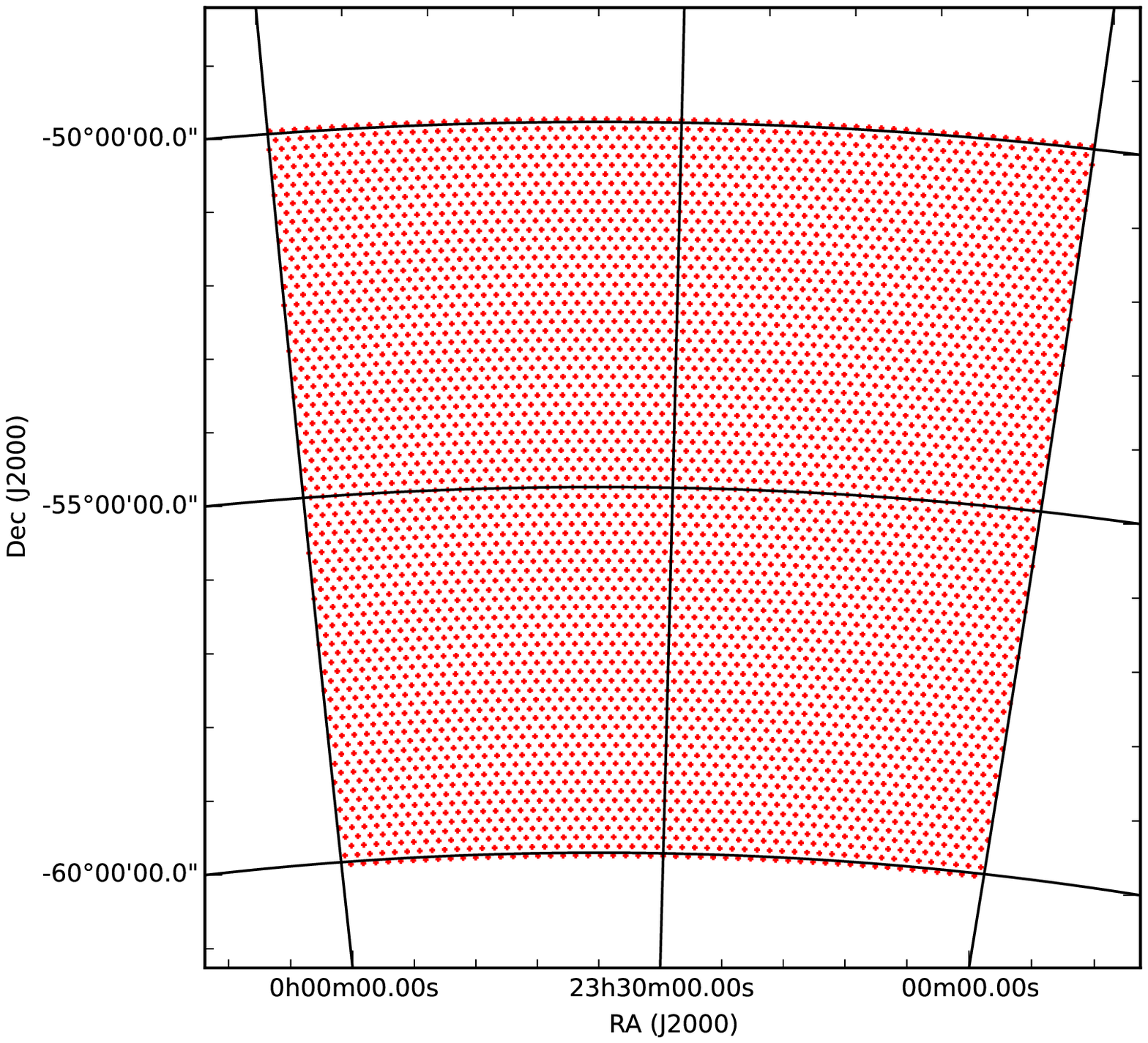}\includegraphics[width=0.5\textwidth]{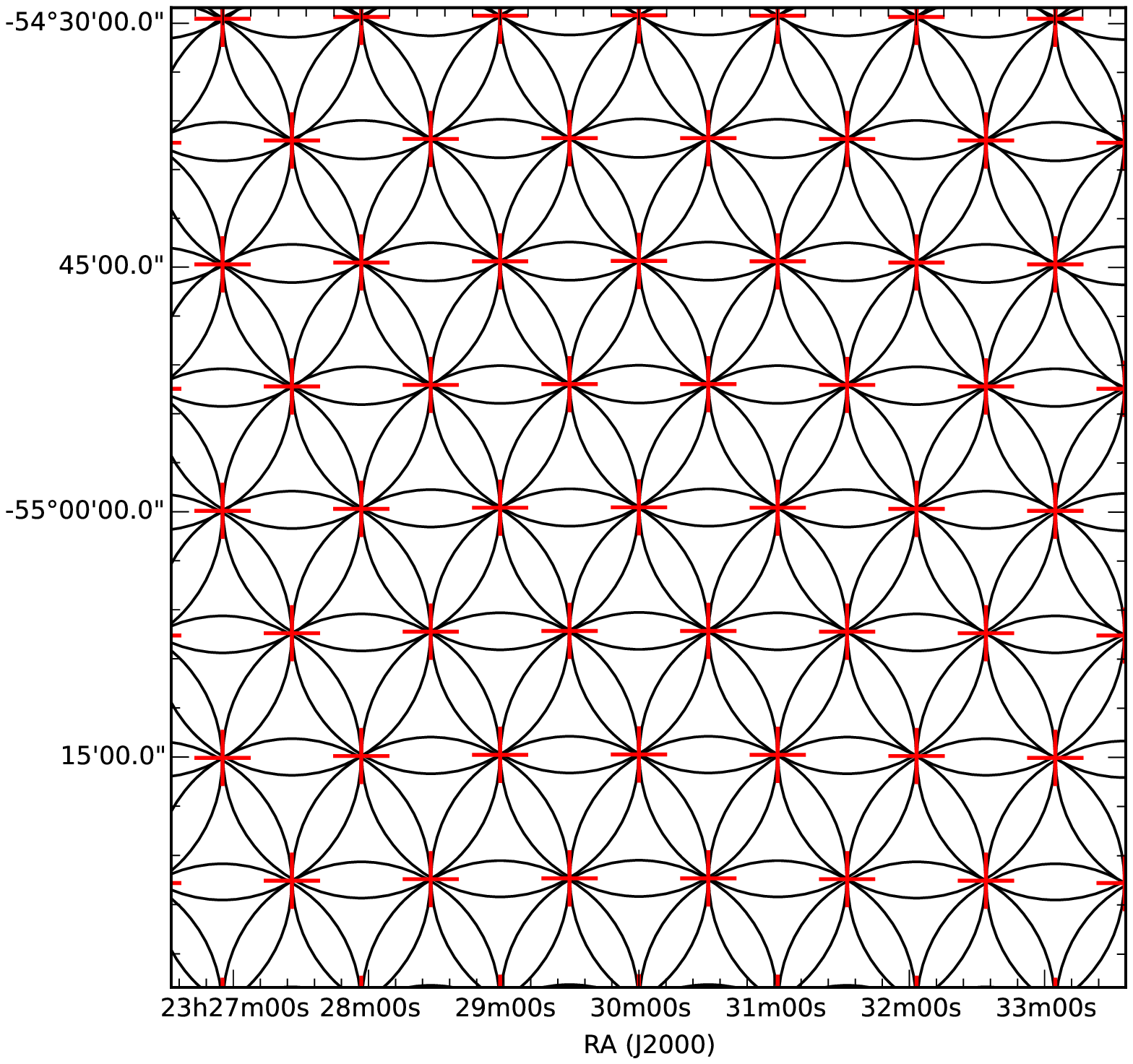}
   \caption{\textit{Left:} Pointing centres of the ATLAS-SPT mosaic. There are a total of 4,787 pointings covering the SSDF between 23h $\leq \alpha \leq$ 24h (0h) and -60\deg $\leq \delta \leq$ -50\deg.
\textit{Right:} A zoomed in section of the pointings at $\alpha=23.5^{\circ},\,\delta=-55^{\circ}$ showing the pointing centres (red +) and the ATCA FWHM at 3.1~GHz. The spacing between pointings has been determined to be at least Nyquist sampling.}
   \label{fig:pointings}
\end{figure}

As the field of view of the ATCA is significantly smaller than the survey area, we conducted the observations as a mosaic of 4,787 pointings, arranged in a hexagonal grid across the field. To ensure all data is retrieved, the spacing between the pointing centres conforms to Nyquist sampling for the high end of the observing band, following $\theta_{\mathrm{hex}} = \frac{\lambda}{D\sqrt{3}}$ where $\theta_{\mathrm{hex}}$ is the hexagonal spacing in radians, $\lambda$ is the observed wavelength and $D$ is the dish diameter. As the ATCA field of view changes rapidly as a function frequency in this band, the low end of the band is significantly oversampled. By spacing the pointing centres using the 3.1~GHz primary beam FWHM we ensure that the spacing is at least Nyquist for all frequencies.  The final pointing spacing is approximately $2.5\times 10^{-3}$ radians scaled along the $\alpha$ axis to $\cos{\delta}$ and is illustrated in Figure~\ref{fig:pointings}. The set of pointings were then divided into groups that would take approximately one hour in sidereal time to complete a single iteration. We refer to these groups of pointings as "blocks".

The 6~km East-West array configurations yield high resolution, but the coverage in the $uv$ plane is quite poor. It is therefore a requirement that each pointing be observed at least 6 times at equidistant hour angles in order to accurately reproduce the morphology of sources while imaging. Considering both the calibration scans and slew times, we were able to integrate on each pointing for 18 seconds per sample of hour angle resulting in a minimum of 108 seconds on each pointing.

\section{Imaging}
Calibration and imaging of the data were carried out using the software package \miriad{} \cite{Sault1995}. We followed standard calibration procedures for wideband imaging as described in the online manual\footnote{http://www.atnf.csiro.au/computing/software/miriad/userguide/userhtml.html}, using PKS 1934-638 as the bandpass and flux calibrator and PKS 2333-528 as the phase calibrator. \ac{RFI} was a major problem for this project as several terrestrial radio sources broadcast within the 1.1-3.1 GHz observing window. Known, persistent sources of \ac{RFI} as well as internal instrumental sources are automatically flagged when new datasets are loaded into \miriad{}. Unpredictable \ac{RFI} was flagged using an automatic flagging task in \miriad{}, \textsc{pgflag} \cite{Offringa2010}. After this process, each mosaic pointing had approximately 25\% of the original visibilities flagged. In addition, approximately 15 hours of observing time was lost to intense wideband \ac{RFI} which saturated the antenna receivers. This time was recovered in a later semester.

After flagging and calibration, each pointing was imaged individually as \miriad{} does not yet support joint deconvolution of wide bandwidth mosaics. The calibrated visibilities were concatenated in time order according to their block number, then split such that each visibility file was of a unique pointing that contained all observed data for that pointing for the project. We then used a customised imaging pipeline to image each pointing individually. 

Using multi-frequency synthesis, we produced dirty maps and beams (i.e. the instrument point-spread function) of each pointing. In addition to the conventional dirty beam, a spectral dirty beam was also produced which enables the spectral information to be preserved in the deconvolution process as described in \cite{Sault1999}. The maps were also forced to be on the same pixel grid by defining a common reference pixel at the field centre.

Due to the sparse $uv$ coverage of each pointing, the instrument response to sources at mJy levels outside of the primary beam caused long linear artefacts above 5$\sigma$. These artefacts could be removed during deconvolution only if the origin source were included in the model. However, given our pixel resolution of 0.86\arcsec{}, deconvolving maps well beyond the size of the primary beam would be extremely computationally expensive. Instead, the pointing was imaged far beyond the primary beam by using a coarser pixel resolution and the approximate positions of strong outlying sources were found using the source-finder \textsc{imsad}. These outlying sources were then imaged and deconvolved as separate "postage stamp" images of 200 pixels square using the full resolution pixel size. The outlier models were then subtracted from the pointing visibilities before re-imaging the primary beam area.

After the outlier subtraction, the dirty maps were deconvolved using the spectral dirty beam to produce clean models which also contain spectral information. Phase self-calibration was performed using the output clean models with the corrections divided into 8 frequency bins which produced far superior calibration results compared with no frequency binning. This self-calibration process was performed 3 times for each pointing, with each iteration cleaning down to deeper flux density thresholds from 20, 10 then finally 5$\sigma$.

The images were produced on the \textit{Galaxy} supercomputer at the Pawsey Supercomputing Centre to drastically reduce the time required to image the field. Since each pointing is imaged independently, they can be processed in parallel. \textit{Galaxy}, with its 472 compute nodes, was able to image hundreds of pointings simultaneously and most imaging was completed unsupervised. Quality checks revealed a limited number of pointings that contained significant image errors, mostly due to strong sources with artefacts that crept into the self-calibration model. These pointings were imaged manually using the same steps as the automatic pipeline but with deconvolution in user-defined regions around sources only.

\begin{figure}
   \centering
   \includegraphics[width=0.8\textwidth]{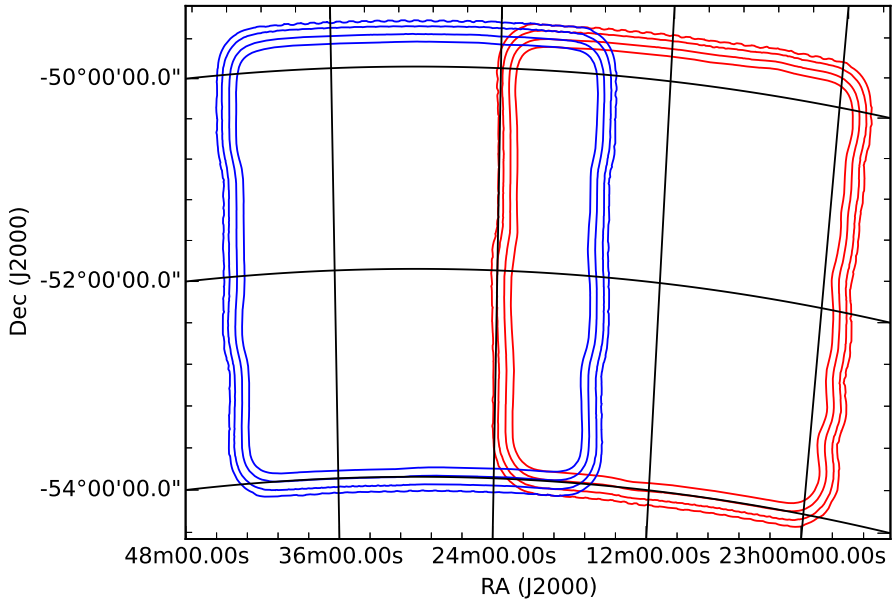}
   \caption{An example of the overlap between two of the 9 mosaics over the covering the ATLAS-SPT field. The contour levels are at 3, 5, 10 and 20$\sigma$ with the two mosaics being distinguished by their contour colour. The reason for the overlap is so that a source in an area of high noise in one mosaic will be in a region of optimal noise in at least one other mosaic.}
   \label{fig:overlap}
\end{figure}

Each mosaic pointing had its clean component model convolved with its respective dirty beam and subtracted from the dirty map to produce a residual map. All models were then convolved with a common Gaussian of 8\arcsec{} (the restoring beam) and added back into their respective residual maps to produce the final restored image for each pointing. The restored images were then primary beam corrected and linear mosaicked. Due to software limitations restricting the number of output pixels, 9 mosaics were produced with each being approximately 20,000 pixels along each coordinate axis and overlapping by one degree. The overlap was to ensure that every region of sky was covered by at least one mosaic with the lowest rms possible; i.e. a source near one of the interior boundaries of the 9 mosaics would be in an area of increased rms due to the lack of overlapping input pointings, however this same source would also appear in the adjacent mosaic with the maximum amount of input pointings contributing to that area of the image resulting in the lowest possible rms. See Fig. \ref{fig:overlap}.

\section{Source Detection and Catalogue}
Our goal is to use automated source detection software due to the number of sources we expect to find. As future radio surveys (i.e. \ac{EMU}) will contain millions of sources, automated techniques will be an absolute necessity and we chose to use this project as an opportunity to gauge the current state of popular source-finding algorithms, building on previous source extraction data challenges \cite{Hopkins2015}. We have experimented with both \textsc{Aegean} \cite{Hancock2012} and \ac{PyBDSM} \cite{Mohan2015} and found that \ac{PyBDSM} better suits our needs of detecting and extracting complex source structures comprising of both compact and extended components. Both \textsc{Aegean} and \ac{PyBDSM} function by calculating a background noise map for an input image, detecting areas of contiguous emissions above a given threshold (called ``islands'') and then fitting multiple Gaussian components to each island. We find that \ac{PyBDSM} is able to produce more accurate source detections with fewer false detections due to the ability to vary the resolution of the background rms map around strong sources likely to have artefacts and the tuneable flagging parameters to exclude unphysical Gaussian components. In an upcoming survey paper, we intend to compare the results produced from \ac{PyBDSM} with those produced by manual source detection and extraction for a smaller region of the field.

At the completion of the project, the output source catalogue will contain a list of sources whose components are grouped together by the island in which they were detected. The full Gaussian component catalogue will also be released. For each source, the source catalogue will include the central sky position, peak flux, integrated flux, background rms, residual rms and source structure classification. A later release will include spectral data.

\section{Bent-Tail Radio Galaxies}
Depending on their orientation, typical radio \ac{AGN} present geometrically symmetrical jets and lobes around the central super-massive black hole. This symmetry is disturbed in the presence of a surrounding moving, dense medium such as the \ac{ICM} found within galaxy clusters which exerts ram pressure onto the jets and lobes which bend them in the direction of movement of the \ac{ICM} producing bent-tail radio galaxies.

Bent-tail radio galaxies are found in large, high-mass galaxy clusters \cite{Blanton2000, Belsole2007, Venturi2007} and are believed to represent radio-loud \ac{AGN} whose jets or lobes are distorted by the moving intra-cluster medium \cite{Mao2010}. Simulations have shown that we can expect to find at least one bent-tail radio galaxy per cluster \cite{Mguda2014}. Our \ac{ATCA} observations are sensitive to \ac{AGN} out to $z \sim 1.3$ and will have plenty of spatial resolution to resolve extended jets and lobes.

\begin{figure}
   \centering
   \includegraphics[width=0.5\textwidth]{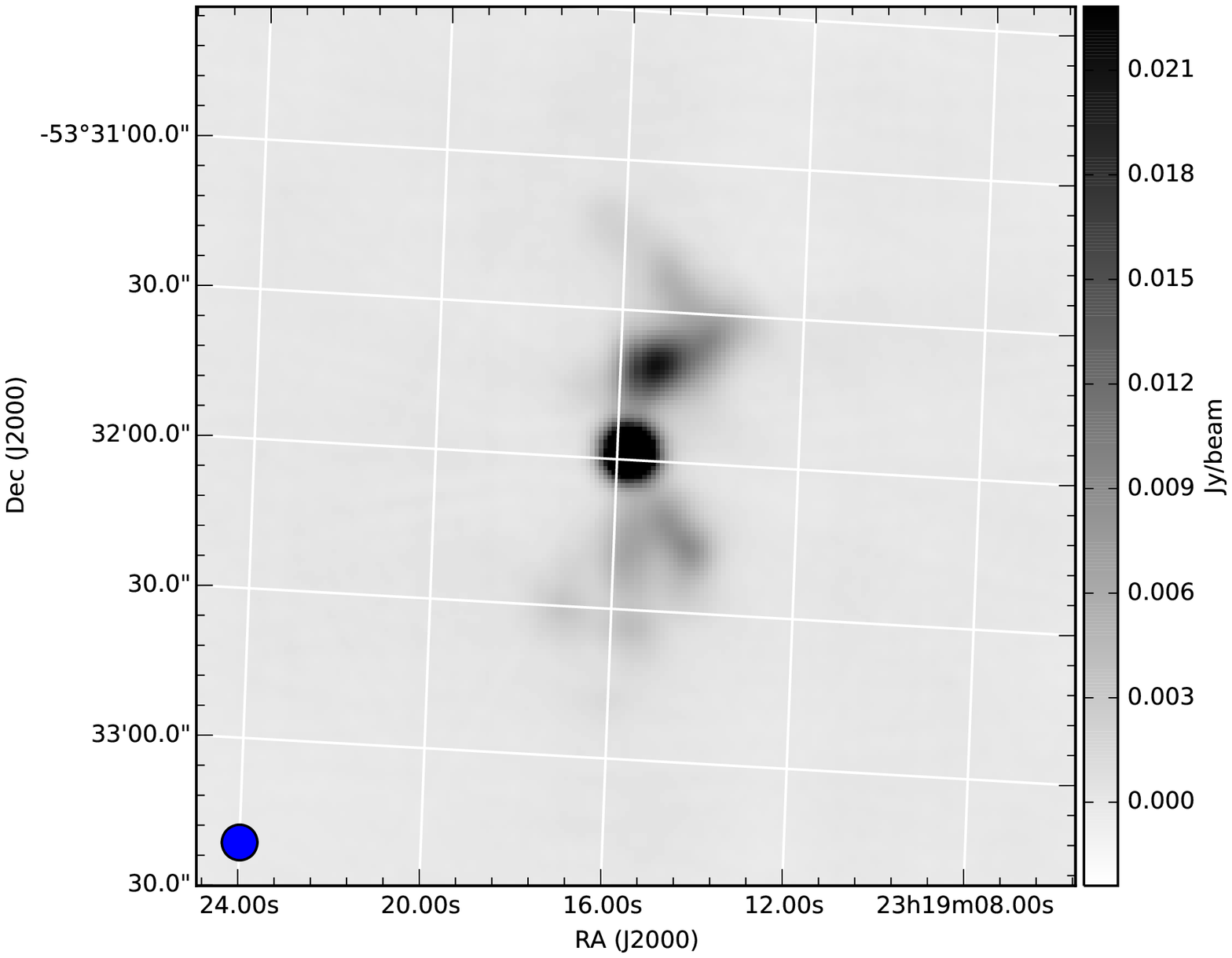}\includegraphics[width=0.5\textwidth]{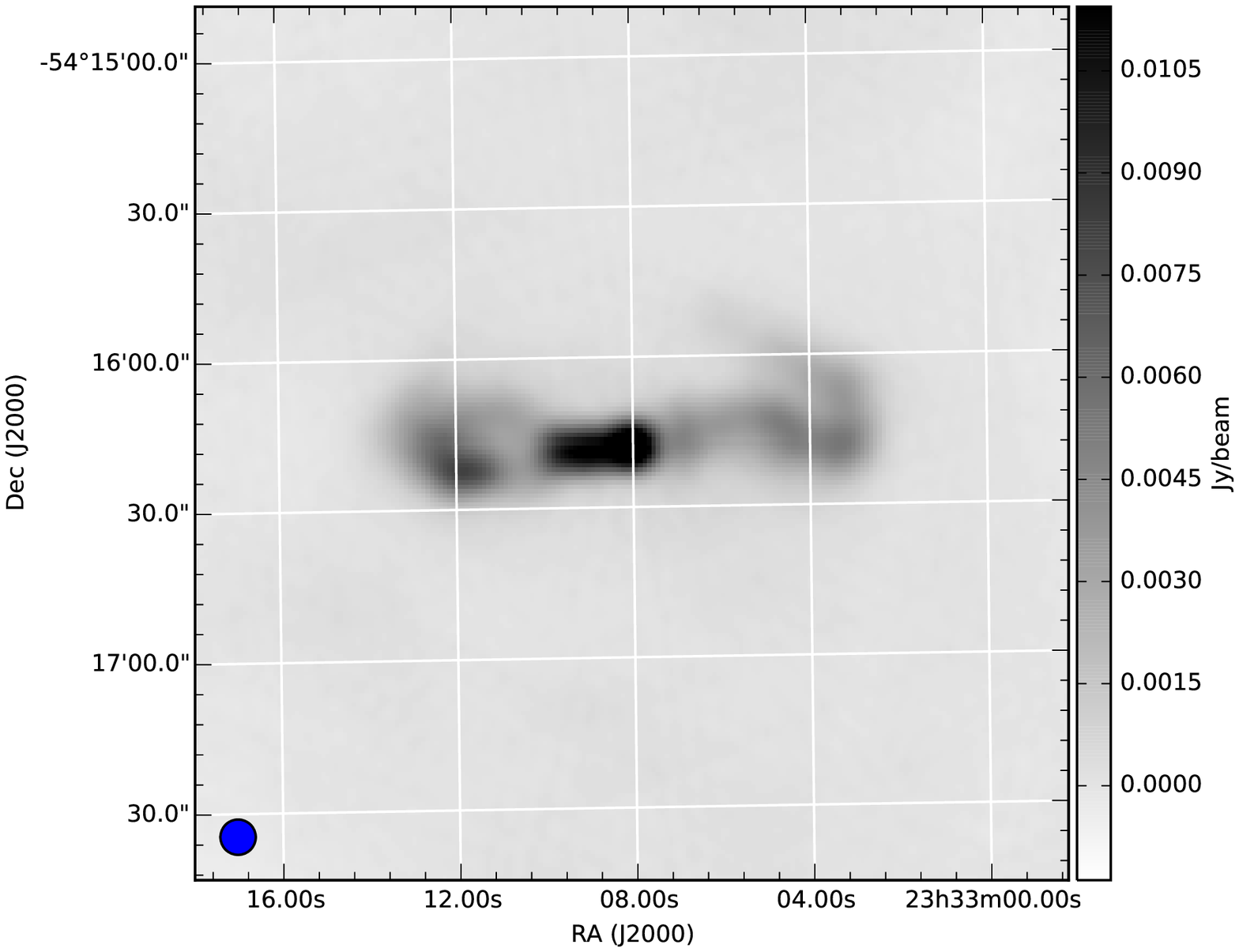}
   \caption{A sample of two bent-tail radio galaxy candidates in the ATLAS-SPT data. The distortion in the jets are due to ram pressure from the moving surrounding \ac{ICM}. The synthesised beam FWHM is shown in the lower left corner of each image.}
   \label{fig:btg-sample}
\end{figure}

We have detected 75 bent-tail radio galaxy candidates in the ATLAS-SPT data, two of which are shown in Fig. \ref{fig:btg-sample}. As the sky density of bent-tail galaxies is low and the determination is not trivial due to the ranges of complex structures, these detections and classifications were made manually. A more quantitative criterion will be published along with the data release in an upcoming survey paper.

\section{Comparison with SZ-Selected Clusters}
Using our bent-tail radio galaxy candidates, we aim to discover new galaxy clusters that were not found with other techniques, in particular with the \ac{SPT} \ac{SZ} survey. We do this by cross-correlating these bent-tail detected clusters with the \ac{SPT} cluster catalogue to further investigate the potential selection effects of \ac{SZ} cluster selection.

While clusters detected by the \ac{SZ} effect should show no redshift bias to at least $z \sim 2$, the \ac{SZ} signal of clusters hosting bright radio sources at high frequencies can be difficult to detect \cite{Lin2007, Martini2009, Lin2009}. It has also been shown that the fraction of \ac{AGN} in clusters is higher for galaxies at higher redshift \cite{Martini2009, Krick2009}, meaning a higher fraction of high-$z$ clusters (compared to low-$z$) may be excluded from \ac{SZ} surveys. Additionally, there is an implied bias toward the most massive clusters in \ac{SZ} surveys since the \ac{SZ} signal is weak and directly related to the cluster mass (similarly for X-ray detected clusters).

Since clusters have a high probability of hosting radio sources \cite{Best2007, VonDerLinden2007} and multi-component radio sources are more often associated with clusters than single component sources \cite{Wing2011}, the technique of using bent-lobe radio galaxies as a method of finding clusters could reveal high-redshift clusters without the requirement for extremely deep observations. These works were based on radio data with a flux density threshold of 1 mJy and optically detected clusters, classifying clusters out to $z \sim 0.5$, whereas our catalogue will extend out to radio sources at $z \sim 1.3$ with an approximate rms of 80\ujybm{} giving a 5$\sigma$ flux density limit of 0.4 mJy.

\begin{figure}
   \centering
   \includegraphics[width=0.75\textwidth]{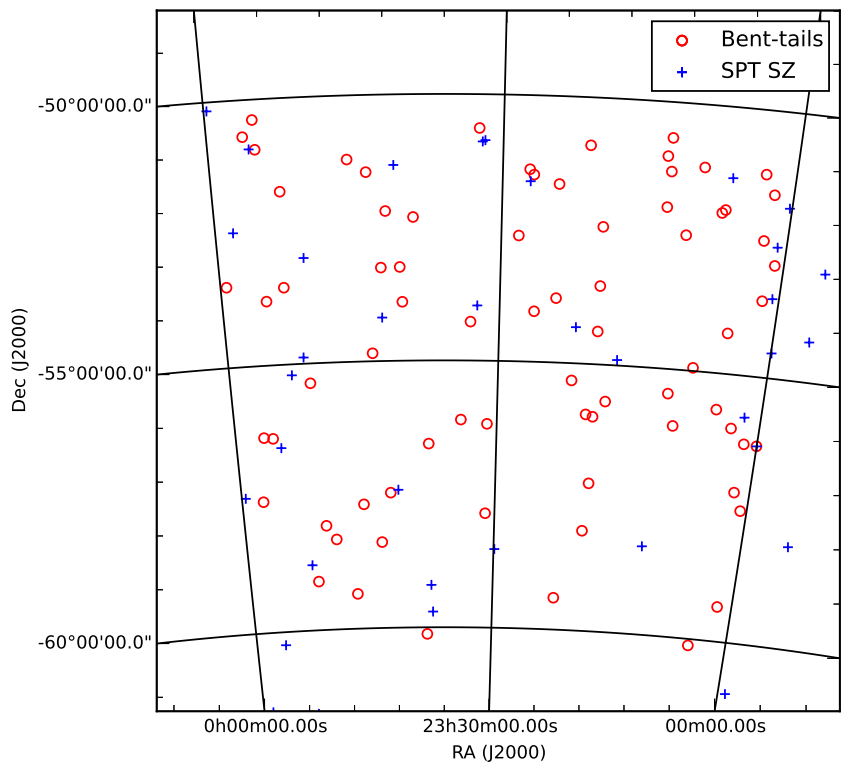}
   \caption{The positions of the bent-tail radio galaxy candidates found in ATLAS-SPT (shown as open red circles) and positions of the galaxy clusters detected in the \ac{SPT} \ac{SZ} survey \cite{Bleem2015} (shown as blue plus markers).}
   \label{fig:btgs}
\end{figure}

Each bent-tail candidate was cross-matched with its nearest neighbour galaxy cluster detected in the \ac{SPT} \ac{SZ} catalogue \cite{Bleem2015}. All of the SZ clusters in the ATLAS-SPT field have either a measured redshift or have a minimum limit presented in the catalogue except for one which we exclude. Since we do not yet have redshifts for our bent-tail candidates, we make the temporary naive assumption that each candidate is associated with its nearest \ac{SPT} \ac{SZ} cluster and we measure the hypothetical distance between the cluster and the bent-tail candidate at the redshift of the cluster. We find that only one bent-tail candidate is within 2 Mpc of its nearest SZ detected cluster, and 4 bent-tails within 4 Mpc. Fig. \ref{fig:btgs} shows the positions of the ATLAS-SPT bent-tail candidates and the \ac{SPT} \ac{SZ} clusters.

\section{Conclusion}
We have reduced and imaged 2.1 GHz radio observations of 100 \sqdeg{} at a resolution of 8\arcsec{} to an rms of 80\ujy{}. Automated source detection and extraction is ongoing and a full catalogue of radio sources will be presented in an upcoming publication. A preliminary analysis into the correlation between galaxy clusters detected in the \ac{SPT} \ac{SZ} survey and bent-tail radio galaxy candidates identified in our ATLAS-SPT image reveals that the bent-tails detected in our data are tracing a different population of galaxy clusters than the \ac{SPT} \ac{SZ} survey. This will be investigated further in our forthcoming survey publication.

\bibliographystyle{JHEP}
\bibliography{bibliography}

\end{document}